\documentclass[10pt]{article}
\usepackage[letterpaper]{geometry}
\usepackage{hicss}
\usepackage{times}
\usepackage[none]{hyphenat}
\usepackage{url}
\usepackage{latexsym}
\usepackage{indentfirst}
\usepackage{graphicx}
\graphicspath{{images/}}

\usepackage{amsmath}
\usepackage{booktabs}
\title{Combinations of Affinity Functions for Different Community Detection Algorithms in Social Networks}

\author{Javier Fumanal-Idocin$^1$, Oscar Cord\'on$^2$, Mar\'ia Min\'arov\'a$^3$, Amparo Alonso-Betanzos$^4$, Humberto Bustince$^1$  \\
  $^1$Public University of Navarra, Pamplona, Spain,
  $^2$Slovak University of Technology in Bratislava, Bratislava, Slovakia \\
  $^3$University of A Coruña, A Coruña. Spain $^4$University of Granada, Granada, Spain \\
  \underline{javier.fumanal@unavarra.es}}

\date{}

\begin{document}
\maketitle
\begin{abstract}
Social network analysis is a popular discipline among the social and behavioural sciences, in which the relationships between different social entities are modelled as a network. One of the most popular problems in social network analysis is finding communities in its network structure.
Usually, a community in a social network is a functional sub-partition of the graph.  However, as the definition of community is somewhat imprecise, many algorithms have been proposed to solve this task, each of them focusing on different social characteristics of the actors and the communities. In this work we propose to use novel combinations of affinity functions, which are designed to capture different social mechanics in the network interactions. We use them to extend already existing community detection algorithms in order to combine the capacity of the affinity functions to model different social interactions than those exploited by the original algorithms.

\end{abstract}

\section{Introduction}

Social network analysis has become an important part of many disciplines such as physics \cite{newman2008physics}, public health and administration \cite{valente2010social, chaves2011social}, and biology \cite{horvath2011weighted}. The idea behind studying complex systems using this technique is that the elements composing such systems are part of multi-lateral interactions and processes, which affect deeply the system and each of its composing individuals. Many problems in computer science have been tackled using social network analysis \cite{newman2018networks}. This is the case of information propagation \cite{lobel2015information}, leader detection \cite{bamakan2019opinion}, recommendation systems \cite{YU2018312, wu2021two} and decision making \cite{zhang2021social, herrera2017consensus}. One of the most popular problems of this kind is community detection.

A community in a social network is a group of nodes that present a significant interaction among them, and much less with the rest of the network. Community detection has been used to identify groups of friends \cite{papadopoulos2012community}, a protein complex \cite{mason2007graph}, and to discover routing strategies \cite{nguyen2014dynamic}. Many algorithms have been proposed to solve community detection, a great deal of them using the concept of modularity to measure how good are the sub-partitions detected \cite{newman2004fast}. The most popular community detection algorithm is the Lovaine algorithm \cite{Blondel_2008}, which uses changes in modularity to obtain the optimal partition. Other popular algorithms that use the idea of modularity are the proposal in \cite{newman2004fast}, which is a greedy algorithm designed to quickly reach a solution, and the proposal in \cite{girvan2002community}, in which the authors propose to iteratively eliminate edges in a network using the idea of edge betweenness. Further approaches include using deep learning to generate an embedding of the graph and a reconstructed adjacency matrix in order to obtain a representation where the community structure can be seen clearly \cite{9097182}; spreading the labels in the network as if it was an epidemic \cite{garza2019community}; or using an equation, called the map equation, to model the information flows of the network \cite{rosvall2009map}.

Another proposal that does not include a modularity optimization process is \cite{fumanal2020community}, in which the authors propose to use a new kind of functions, the affinity functions, that model the local interactions between actors according to different social behaviours. They do so in order to propose a community detection algorithm, the Borgia Clustering, that simulates the dynamics of the conquests of Cesare Borgia in the Italian Renaissance. However, the possibilities of using affinity functions with other community detection algorithms were not studied. This possibility is of spatial relevance because affinity functions could improve the network features that some community detection algorithms exploit. In fact, some algorithms compute different representations of the network using computationally expensive processes, like a convolution \cite{9097182}. Affinity functions could work in a similar way in those algorithm that do not compute other representations for the original network.  
Besides, although some combination of affinity functions were proposed in \cite{fumanal2020community}, some mathematical properties of this combination could be relaxed in order to explore further results.

Taking these considerations into account, in this work we aim to:
\begin{enumerate}
    \item Extend the convex combination of affinity functions to a more general expression.
    \item Explore the possibilities of using affinity functions with community detection algorithms other than the Borgia Clustering.
\end{enumerate}

In order to achieve our aims, we use different combinations of affinity functions without restricting them to be convex combinations, and then we study these combinations with three of the most popular modularity-based community detection algorithms: the Lovaine algorithm \cite{Blondel_2008}, the Greedy modularity algorithm \cite{newman2004fast} and Girvan-Newman algorithm the \cite{girvan2002community}, and compare the results with the Borgia Clustering algorithm.

The rest of the work follows this structure: in Section~\ref{sec:mthods} we describe the affinity functions used and how we combine them, and we also discuss the community detection algorithms used in our experimentation. In Section~\ref{sec:exp} we discuss the datasets used in our experimentation and the results obtained. Finally, in Section~\ref{sec:res} we draw our conclusions for this work and the future lines for our research.

\section{Methods} \label{sec:mthods}
In this section we recall the notion of affinity function, the functions studied and the algorithms tested to perform community detection.

\subsection{Affinity functions}
Affinity functions were proposed in \cite{fumanal2020community} to measure the strength of the relationship between a pair of actors in a social network by capturing the nature of their local interactions. They are defined as functions that take as input two different actors, $x$ and $y$, and return a number in the $[0,1]$ interval: 
\[
F_C:(x, y) \rightarrow [0,1]
\]

where $C$ is the adjacency matrix whose each entry $C(x, y)$ quantifies the relationship for the pair of actors $x, y$ in a weighted network. The affinity value will be higher or lower depending on which aspect of the relationship we are taking into account, like the number of friends in common or the relative strength of the interaction between $x$ and $y$. Some affinity functions are:

\begin{itemize}
	\item \textit{Best friend affinity}: it measures 
	the importance of a relationship with an actor $y$ for the actor $x$, in relation to all the other relationships of $x$: 
	\begin{equation} \label{eq-bf}
		F^{BF}_C (x,y)= \frac{C(x,y)}{\sum_{a \in V} C(x,a)}
	\end{equation}  
	
	\item \textit{Best common friend affinity}: it measures the importance of the relationship taking into account how important are the common connections between the connected nodes to $x$ and $y$, in relation to all other relationships of $x$ in the network:
	\begin{equation} \label{eq-bcf}
		F^{BCF}_C (x,y)= \frac{\max_{a \in V }\{\min \{ C(x,a), C(y,a)\} \} }{\sum_{a \in V} C(x,a)} 
	\end{equation}
	
	\item \textit{Machiavelli affinity}: it computes how affine two actors $x$ and $y$ are based on how similar is the social structure that surrounds them:
	\begin{equation} \label{eq-mac}
	F^{Mach}_C (x,y)= 1 - \frac{\mid I_x - I_y \mid}{\max \{I_x, I_y\} }, 
	\end{equation} where $I_a = \sum_{z \in Z} D(z)$, where $Z$ is the set of actors where $C(a, z) > 0, \forall z \in Z$, and $D(z)$ is the centrality degree of $z$. 
\end{itemize} 

Regarding the interpretation of the affinity value obtained, a $0$ value means that no affinity has been found at all while an $1$ value means that there is a perfect match according to the analyzed factors
Since affinities are not necessarily symmetrical, the strength of this interaction depends on who the sender and receiver are, as it happens in human interactions. We denote a network resulting from calculating all its edges with an affinity function an ``affinity network".

There are two different kinds of affinity functions: personal and structural affinities. Personal affinities are computed using the edges of the $x, y$ actors, while the structural affinities use different centrality measures of these actors. Two examples of personal affinity functions are the best friend and best common friend affinities. The Machiavelli affinity is an example of a structural affinity. 

Depending on the affinity function computed, the resulting affinity network will have different structural properties. For example, for the case of the ratio of connections in the network with the respect to all the potential ones (commonly called density): the Machiavelli affinity will result in a network with density equal to $1$; the best friend affinity will preserve the same density as in the adjacency network; and the best common friend affinity tends to increase the density with respect to the adjacency network.

More affinity functions and further details about them can be found in \cite{fumanal2020community}.

\subsection{Combinations of affinity functions}

The convex combination of two affinity functions is also an affinity function. This proof is straightforward: the result of an affinity function is a number in the $[0,1]$ interval. The convex combination of two numbers in the $[0,1]$ interval is another number in the same interval. So the operation that computes the convex combination of two affinity functions is also a valid affinity function.

However, some combinations of affinity functions that are not convex are also affinity functions. As long as we guarantee that the output is inside the $[0,1]$ interval, the resulting value is a valid affinity function. We are interested in these expressions because some affinity functions tend to have a higher average value than others. Combining affinity functions can be problematic because the values of one of the functions can be irrelevant compared to the other in most cases. One possible solution to this problem is to extremely skew the value of the $\alpha$ to favour the irrelevant affinity. However, this solution presents two problems:

\begin{itemize}
    \item This solution results not only in decreasing one affinity, it also increases the other, which is not an indented behaviour. Also, finding the proper equilibrium between the mixing parameters can require an expensive fine-tuning.
    \item The interpretability of this solution is counter-intuitive. If one of the mixing parameters is much higher than the other, we can interpret that one affinity function is much important than the other in our analysis. However, this would not be the case, as we did not increase on purpose that parameter, we only tried to decrease the average value of the other affinity function.
\end{itemize}

Using a non-convex combination, might solve both of these problems. By simply reducing one mixing parameter without affecting the other, we are able to scale both affinity functions to an appropriate average value without performing an expensive fine-tuning or affecting the interpretability of the combination.

To do so, we start by substituting the expression of a convex combination of two affinity functions:

\begin{equation} \label{eq:cc}
    F^{CC}_C = \alpha F^{BF}_C (x,y) + (1 - \alpha) F^{BCF}_C (x,y)
\end{equation}

with $\alpha \in [0,1]$, for the linear combination:

\begin{equation} \label{eq:ncc}
    F^{LC}_C = \alpha F^{BF}_C (x,y) + \beta F^{BCF}_C (x,y)
\end{equation} 
with $\alpha, \beta \in [0,1]$.

However Eq. (\ref{eq:cc}) might not be always an affinity function, so we establish the constraint that: 
\begin{equation}\label{eq:ab}
    \alpha + \beta <= 1  
\end{equation}
Of course, if $\alpha + \beta = 0$, then we obtain a matrix of zeros as result. Although that result it is technically a valid affinity function, we also add another constraint:
\begin{equation}\label{eq:ab2}
    \alpha + \beta > 0  
\end{equation}

In this way, if $\alpha + \beta = 1$ we recover a convex combination, and if not, we are obtaining a value lesser than the one obtained using a convex combination, so it will never be above $1$. We denote the expression Eq. (\ref{eq:ncc}), as long as it holds Eq. (\ref{eq:ab}) and Eq. (\ref{eq:ab2}), a less than-convex combination.

It is straightforward that the less than-convex combination of two affinity functions is another valid affinity function, because it will always be lower than the convex combination of affinity functions (which is a valid affinity function), and always bigger than $0$.

\subsection{Community detection algorithms}
For our experimentation we have studied three very popular community detection algorithms based on different modularity optimization processes:
\begin{itemize}
    \item \textit{Lovaine community detection algorithm} \cite{Blondel_2008}: this method optimizes the modularity of the community structure found as the algorithm progresses. The algorithm is divided in two phases:
    (1) First, small communities are found by optimizing modularity locally on each node. We do so by computing the changes in modularity when each node $i$ is joined in the same community as each of their neighbours, and choosing the biggest positive change in modularity for each node. We do so for each node until no positive change in modularity is possible. (2) Then, each of these communities is grouped into one node and the process repeats iteratively.
    \item \textit{Girvan-Newman community detection algorithm} \cite{girvan2002community}: this algorithm progressively removes edges from the original network based on the idea of how likely an edge is to be a bridge between communities, which is called ``edge betweenness''. The algorithm repeats iteratively this process  until there are no more edges: (1) We compute all the edge betweenness. (2) We remove the one with the highest value. Finally, we obtain a dendrogram where the leaves are the individual nodes.
    \item \textit{Greedy modularity community detection algorithm} \cite{newman2004fast}: is a greedy algorithm designed to reach a solution in few steps. It works by optimizing the modularity value of each partition in the graph. Each node starts as the sole member of its community. Then, we select the edge from the original graph that has the highest positive modularity change. We do so iteratively, until only all the nodes are connected. Then, we choose the partition with better modularity as the final result. 
\end{itemize}

We chose these algorithms as they are very popular, and the concept of modularity is one of the most popular ones in community detection in social network analysis. The procedure to use them with affinity functions  is the same for the three algorithms: we compute the affinity function or the combination of affinity functions on the original adjacency matrix, and then, we compute the algorithm on the resulting affinity network.

For the sake of comparison, we have also compared the results obtained with these algorithms with those obtained by the Borgia Clustering \cite{fumanal2020community}. The Borgia Clustering is a community detection algorithm that generalizes the gravitational clustering algorithm \cite{Wright} using affinity functions, among other changes. This algorithm was explicitly designed to work with affinity functions, so we think that it is interesting to compare the results obtained using this algorithm with the rest of the community detection algorithms using affinity functions.

\section{Experimental results} \label{sec:exp}

In this section we have compared the results performing community detection in three datasets:
\begin{itemize}
    \item Zachary Karate Club \cite{zachary1977information}: is a social network of a university karate club that consists of 34 people and how they split into two communities.
    
    \item Dolphin \cite{lusseau2003bottlenose}: a social network of bottlenose dolphins, where each link represents the frequency in which they played.
    
    \item Polbooks \cite{adamic2005political}: nodes represent books about US politics sold by Amazon. Edges represent frequent co-purchasing of books by the same buyers, as indicated by the "customers who bought this book also bought these other books" feature on Amazon.
\end{itemize}

\subsection{Performance metrics}

\subsubsection{Normalized Mutual Information}
Normalized Mutual Information (NMI) is a normalization of the Mutual Information score to scale the results between 0 (no mutual information) and 1 (perfect correlation). We have used the NMI to evaluate the results obtained in community detection against ground truth labels.

Given the conditional entropy of two populations:
\begin{equation}
\begin{split}
H\left(Y|X\right)&=-\sum_{x}\sum_{y}p\left(x,y\right)\log p\left(y|x\right)
\end{split}
\end{equation}
then, the mutual information of two random discrete variables can be defined as:\\
\begin{equation}
I(X;Y) = H(X)-H(X|Y),
\end{equation}
which measures the mutual dependence between $X$ and $Y$. So, the final expression for the NMI is:
\begin{equation}
NMI\left(X,Y\right)=\dfrac{2I\left(X;Y\right)}{H\left(X\right)+H\left(Y\right)}
\end{equation}

\subsubsection{Modularity}

Modularity is a metric of the structure of networks that measures the strength of partition into different communities  \cite{newman2004finding}. Networks with high modularity have dense connections between the nodes within communities but scarce connections between nodes in different ones.


The expression of modularity, $Q$, is the following:
\begin{equation}
    Q = \sum_{i=1}^c (e_{ii} - a^{2}_i )
\end{equation}
where $c$ is the number of communities, where $e_{u,v}$ is the fraction of edge with one end in a community, $u$, and another end in community $v$, and $a$ is the fraction of edges with one end in the $i$-th community.

\subsection{The effects of different affinity combinations in the networks}
In this subsection we have tested the effects in each dataset using different affinity functions compared to the original adjacency matrix. We expect the best friend affinity not to alter the degree in each node, but other centrality measures may be affected. And for the best common friend affinity, we expect it to increase the degree in each node, which might reduce other centrality measures, specially the betweenness, because more edges will be present in the network.

\begin{figure}[ht]
    \centering
    \begin{tabular}{c}
    \includegraphics[width=0.5\linewidth]{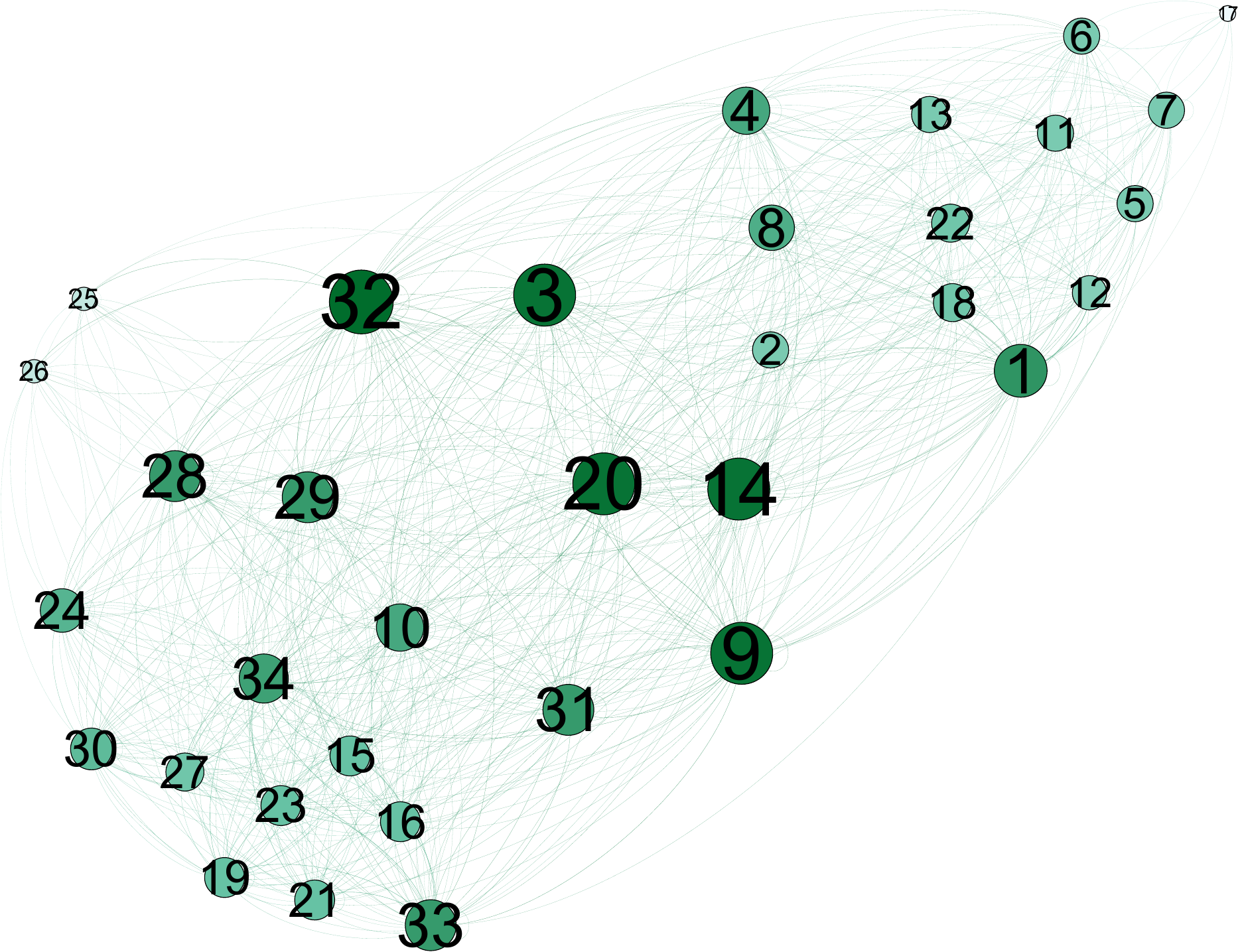}\\
    a) $\alpha=0.25, \beta=0.50$\\
    \includegraphics[width=0.5\linewidth]{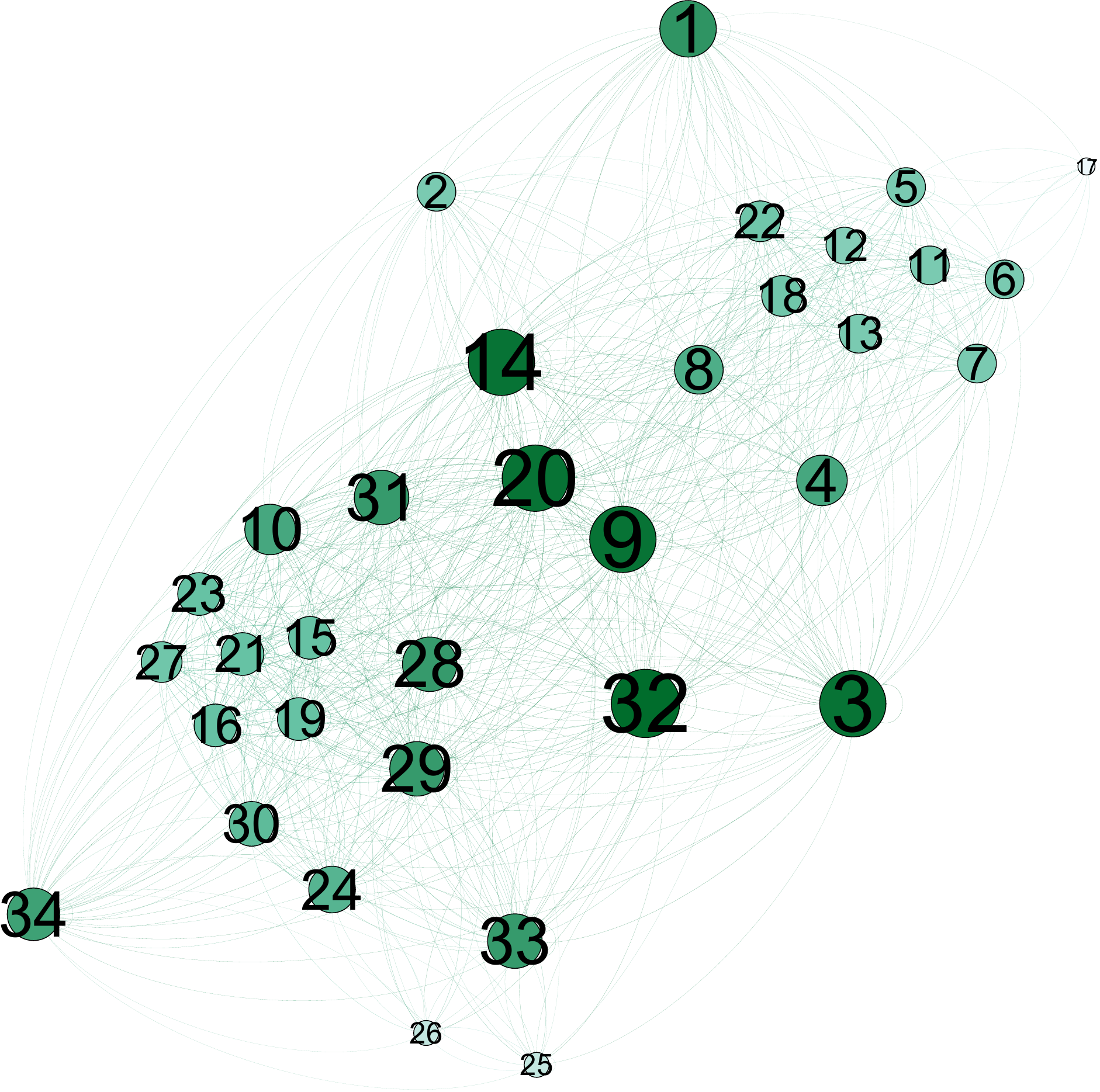}\\
    b) $\alpha=0.50, \beta=0.10$\\
    \includegraphics[width=0.5\linewidth]{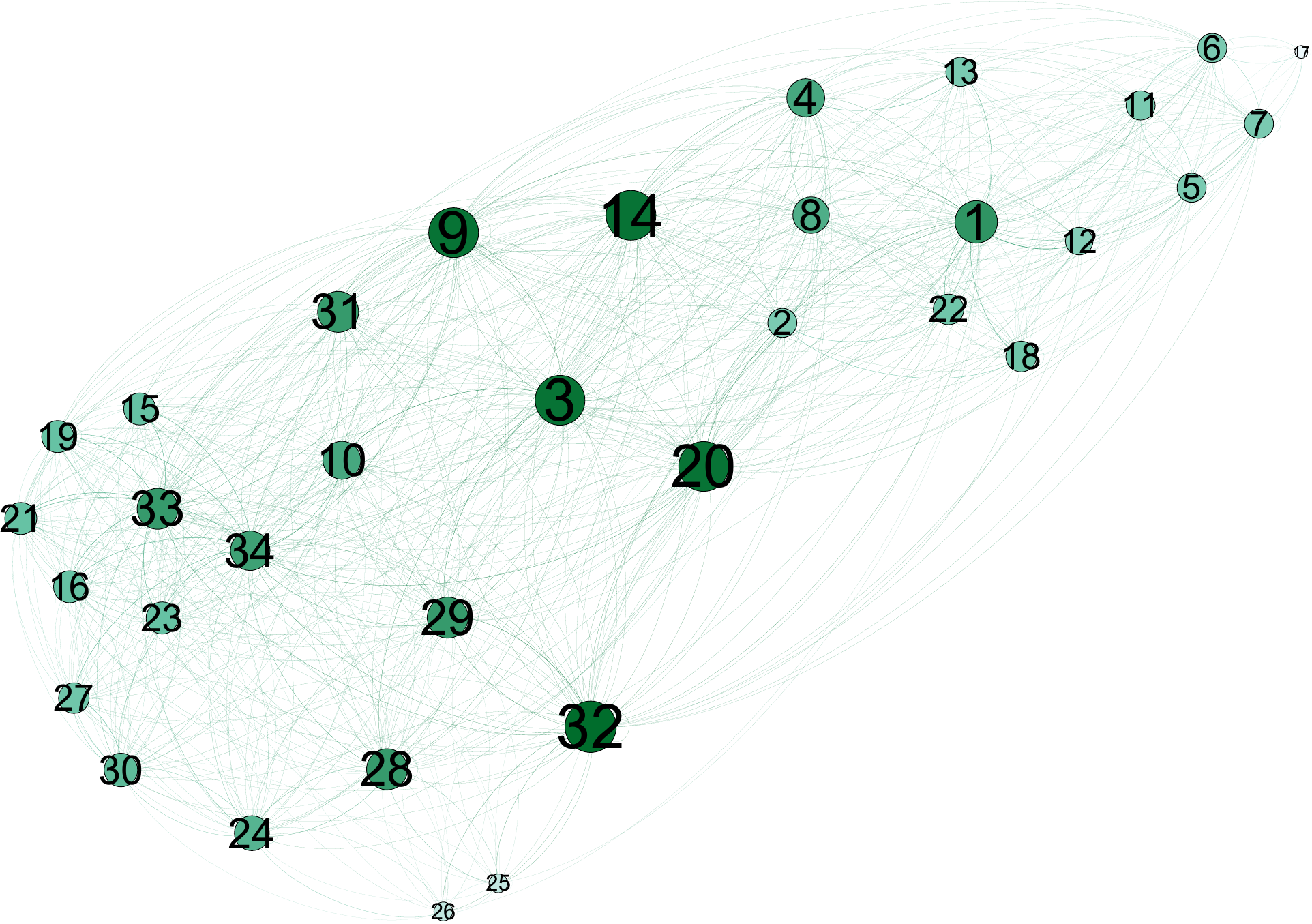}\\
    c) $\alpha=0.75, \beta=0.25$
    \end{tabular}
    \caption{Visual representation of the Zachary Karate club Network using three different combinations of affinities, using the parameters indicated for each one. We can observe that two communities are clearly present in the three networks, although the frontier between both is more clearly present in \textbf{(b)} compared to \textbf{(a)} and \textbf{(c)}. }
    \label{fig:nets_af_zac}
\end{figure}

An example of three different affinity functions applied to the Zachary karate network is shown in Figure \ref{fig:nets_af_zac}. We can see that adding the best common friend affinity did increase the number of edges and the network, and the role that some actors present in the network varies significantly depending on the affinity function used. We expect the results in community detection to change considerably since the centrality measures of some actors change significantly. Specially relevant are the changes of actors ``$34$'' or ``$2$''. It is also noteworthy how some actors do not change their role in their network independently of the affinity used, like ``$17$'' or ``$8$''.  
In this particular case, we can see that even though there are clearly two communities in three affinity networks, the border between them varies significantly between them since some actors in between the communities seem to differ their position remarkably in the different affinity networks.

\subsection{Community detection experiments}
In this subsection we have evaluated the results obtained using the combinations of affinity functions with community detection algorithms other than the Borgia Clustering, and then we have compared them with those obtained using this algorithm. We use the Modularity as internal index and the NMI as an external one to evaluate our results. 

In Table \ref{tab:classic} we have displayed the results obtained for each dataset and algorithm using the original adjacency matrix. We can see that the Greedy Modularity algorithm obtained the best average results in the NMI index and the Lovaine algorithm did so for the Modularity index.

\begin{figure*}[ht]
    \centering
    \begin{tabular}{cccc}
    \toprule
        & Greedy Modularity & Girvan-Newman & Louvain \\
        \midrule
         Zachary     & \includegraphics[width=0.3\linewidth]{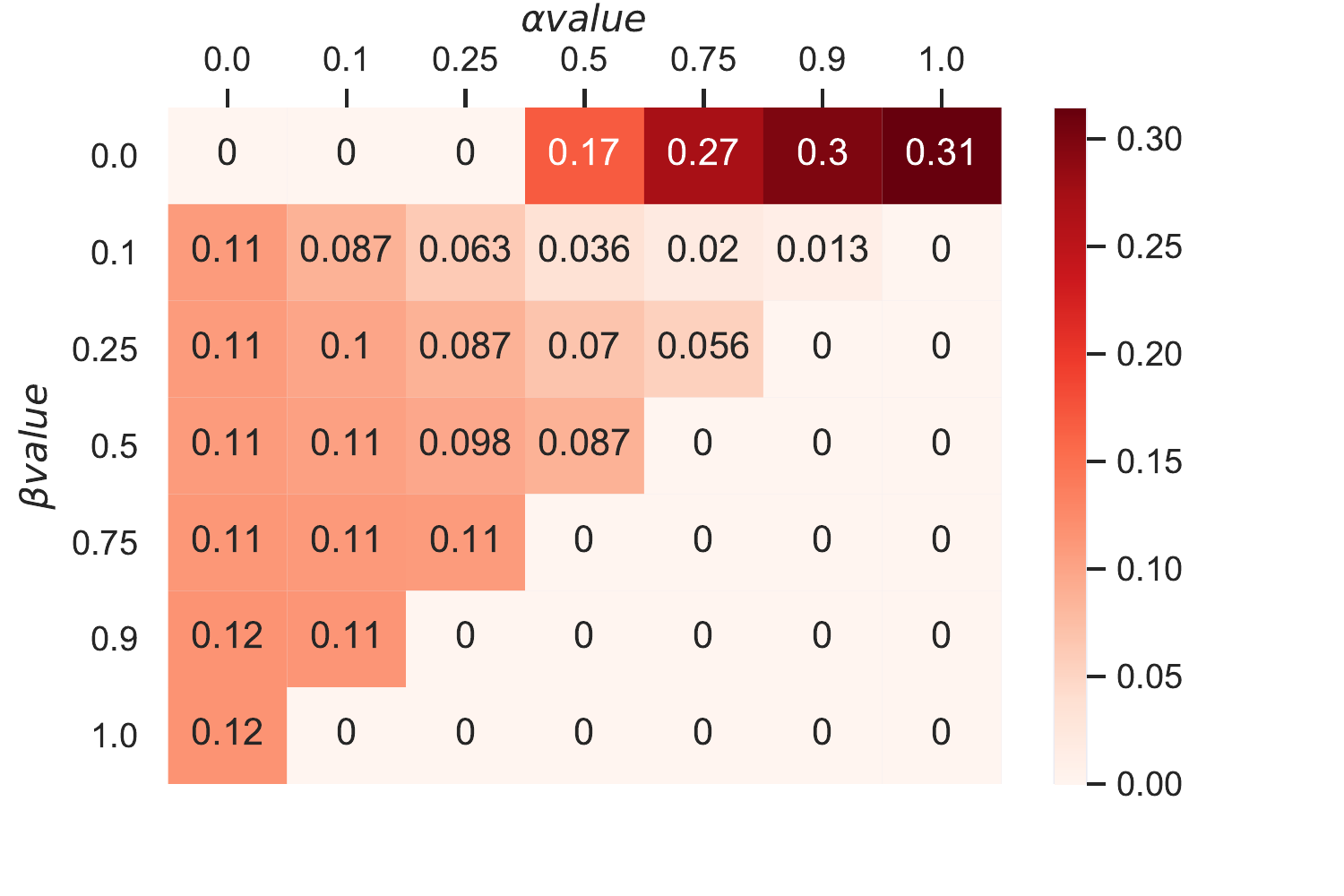} & \includegraphics[width=0.3\linewidth]{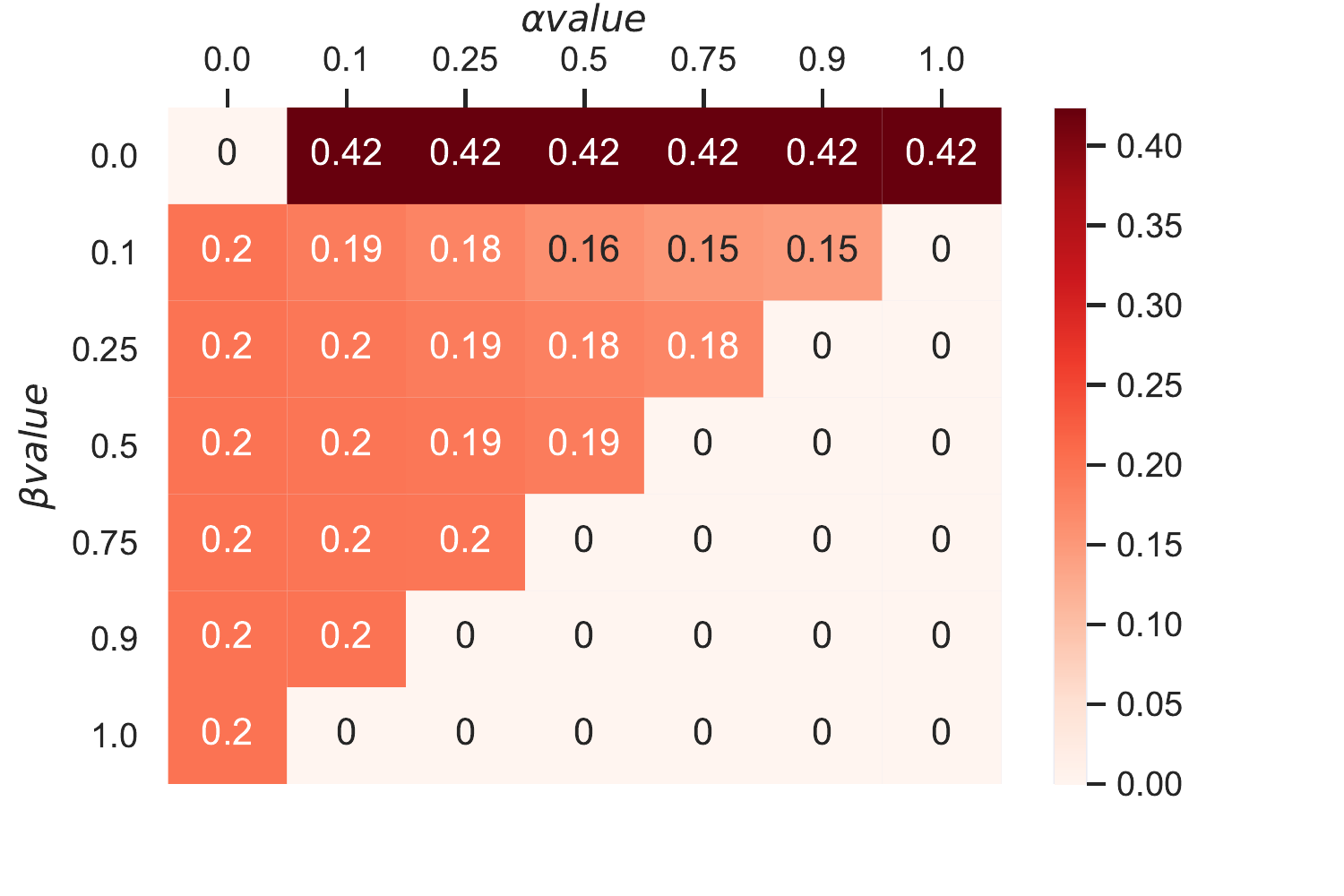} &
         \includegraphics[width=0.3\linewidth]{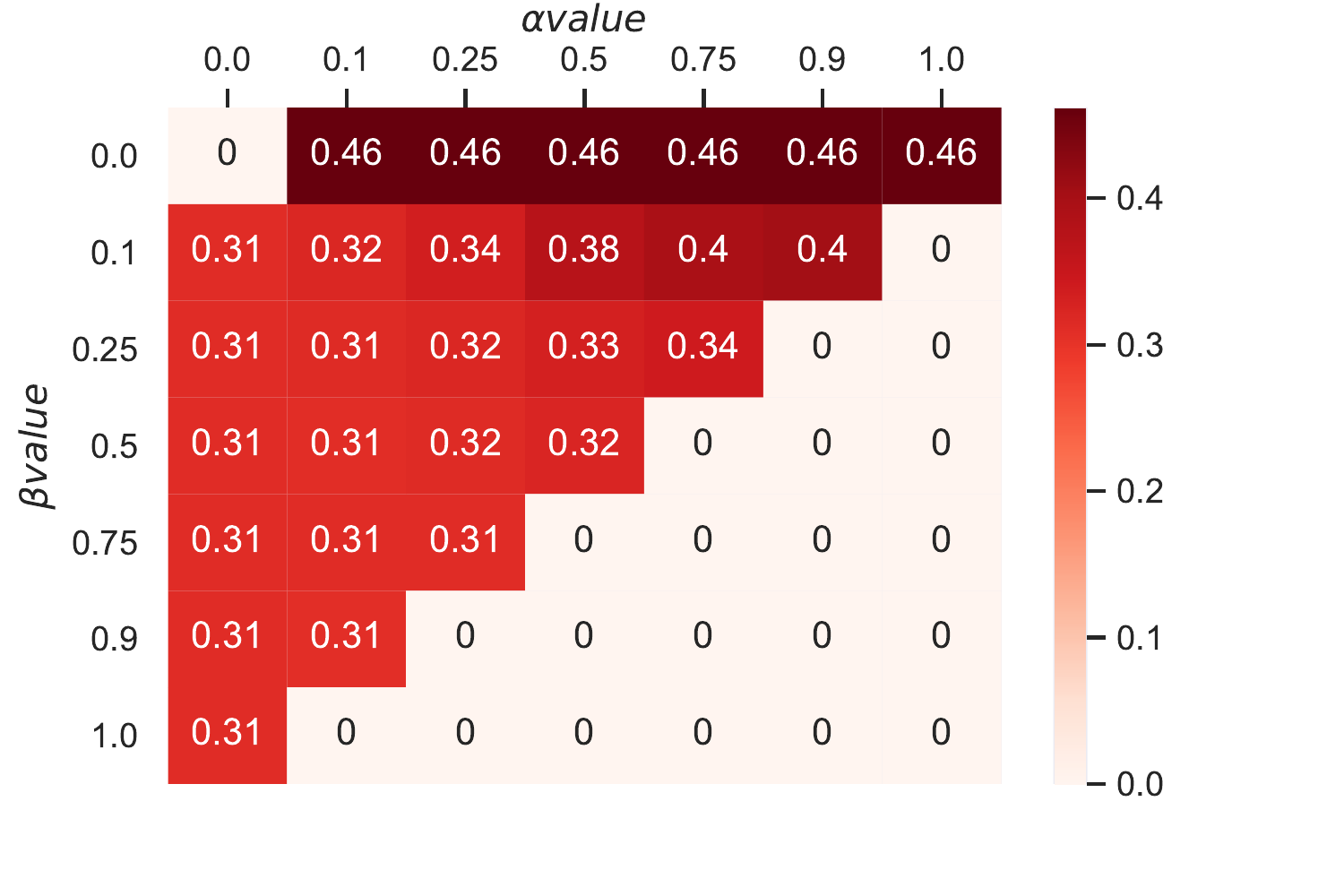}  \\
         \midrule
         Dolphin & \includegraphics[width=0.3\linewidth]{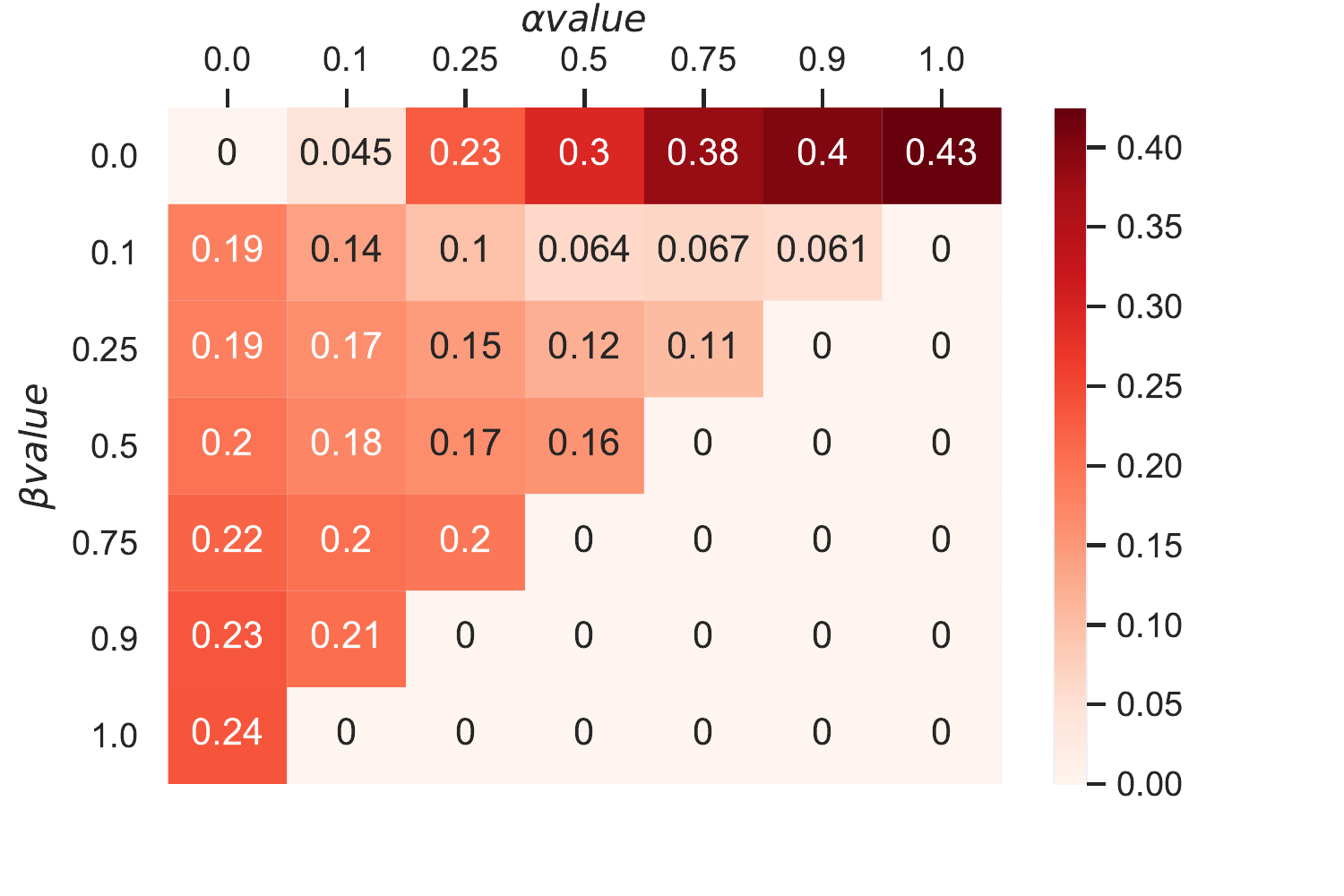} & \includegraphics[width=0.3\linewidth]{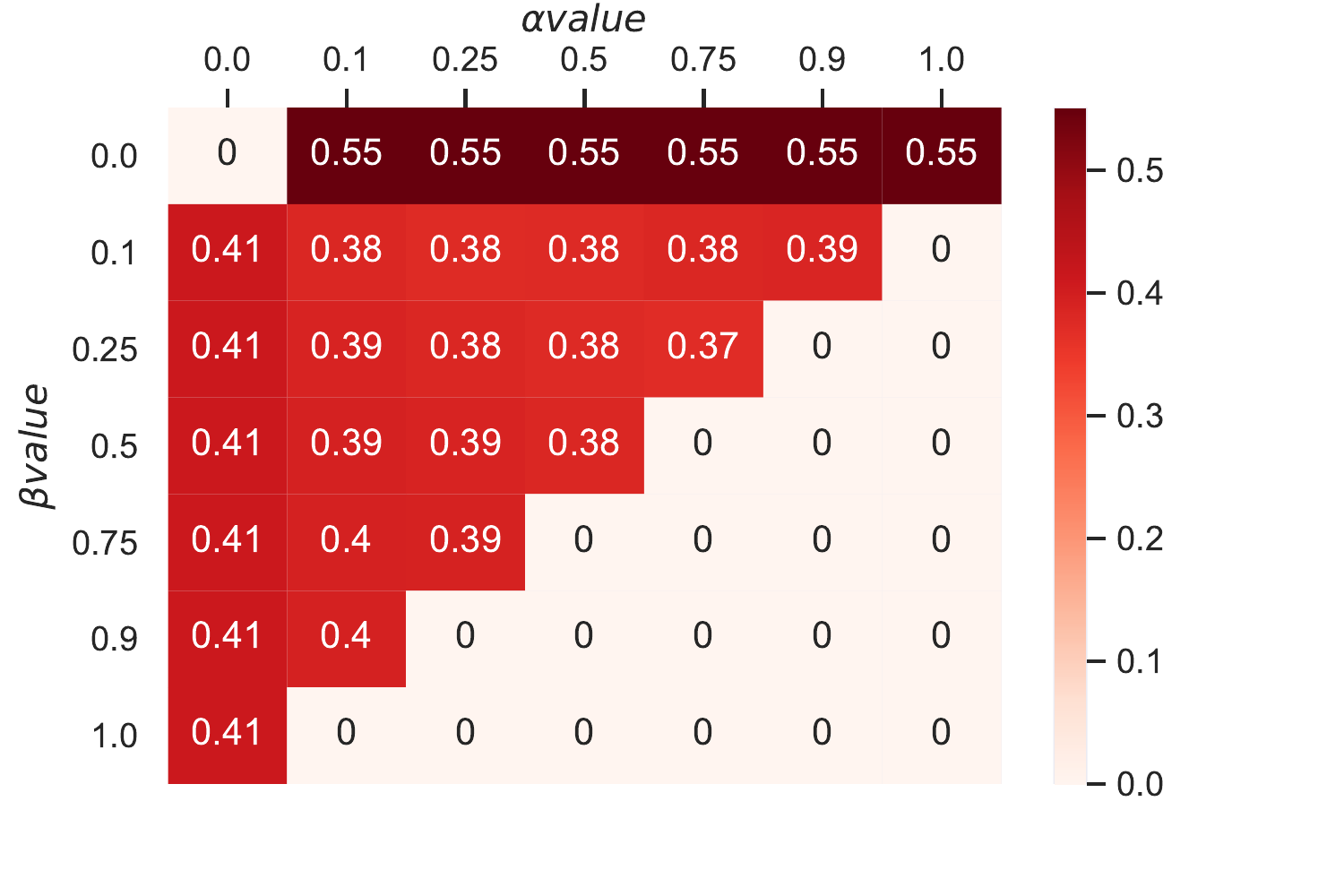} & \includegraphics[width=0.3\linewidth]{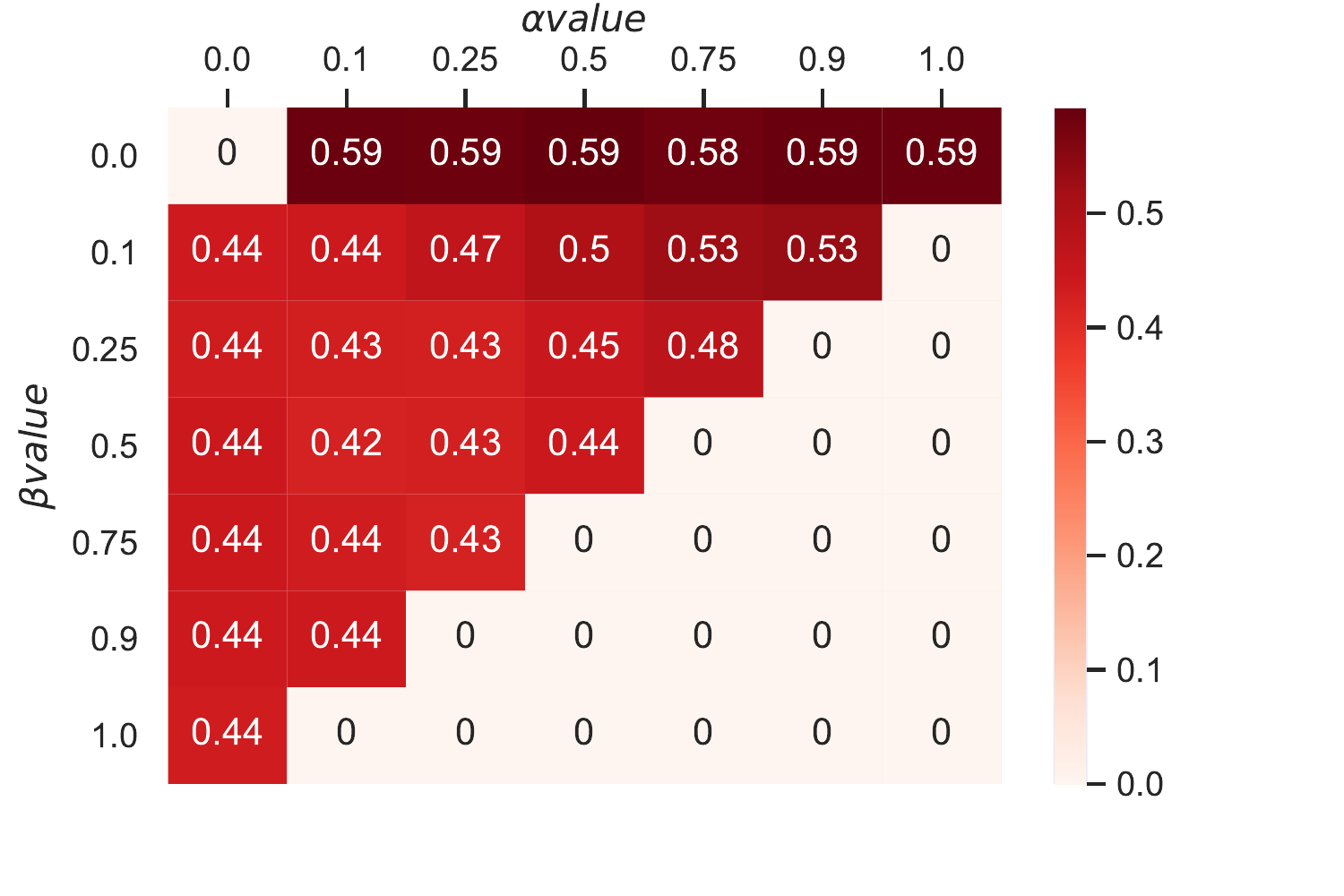}  \\
         \midrule
         Polbooks & \includegraphics[width=0.3\linewidth]{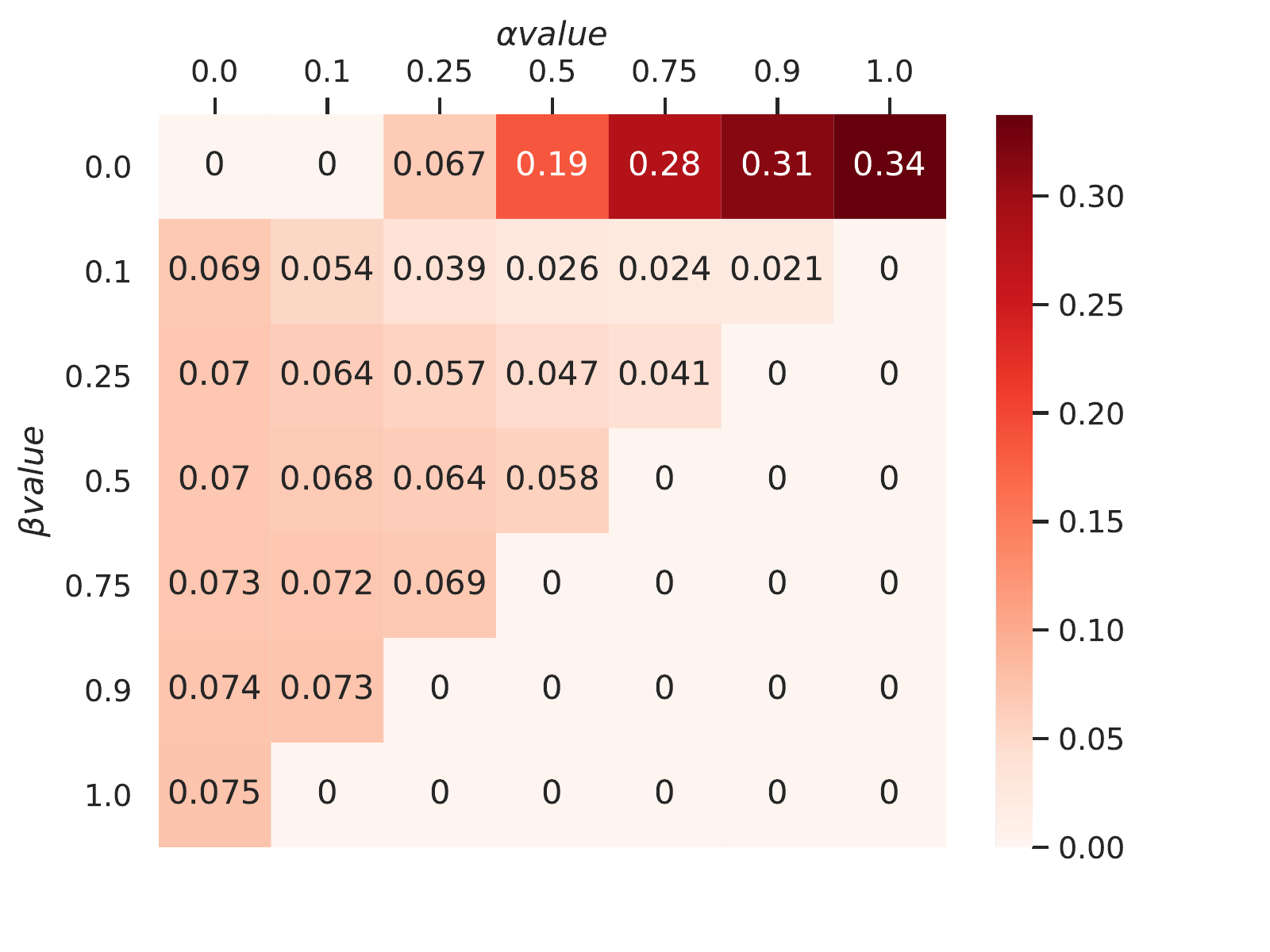} & \includegraphics[width=0.3\linewidth]{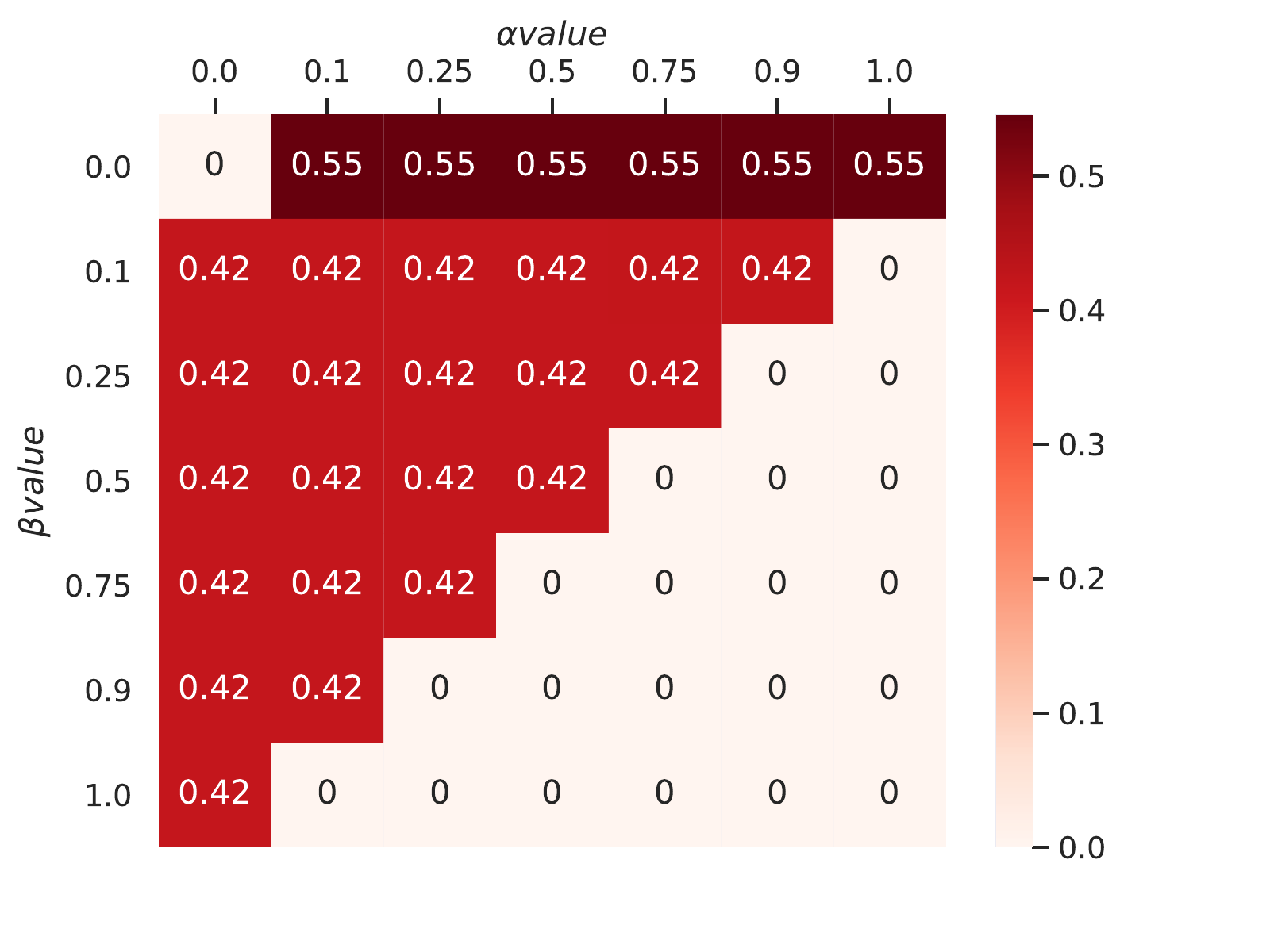} & \includegraphics[width=0.3\linewidth]{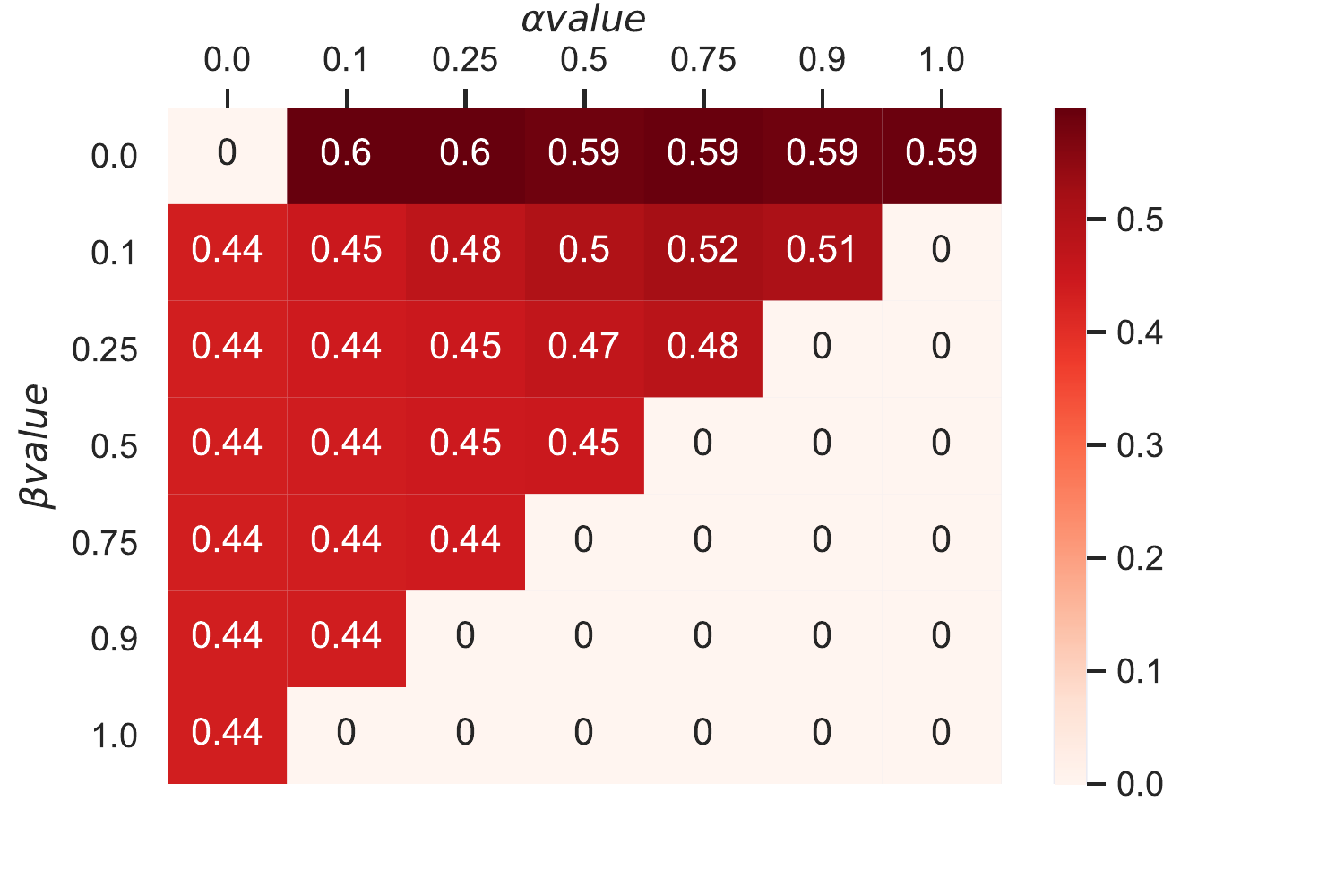}  \\
         \bottomrule
    \end{tabular}
    \caption{Modularity results for the community detection algorithms using affinity networks. Best friend affinity overperforms the combination with the best friend affinity in all cases. Results are generally better than those obtained with the adjacency matrix according to this metric.}
    \label{fig:mod}
\end{figure*}

\begin{figure*}[ht]
    \centering
    \begin{tabular}{cccc}
    \toprule
        & Greedy Modularity & Girvan-Newman & Louvain \\
        \midrule
         Zachary     & \includegraphics[width=0.3\linewidth]{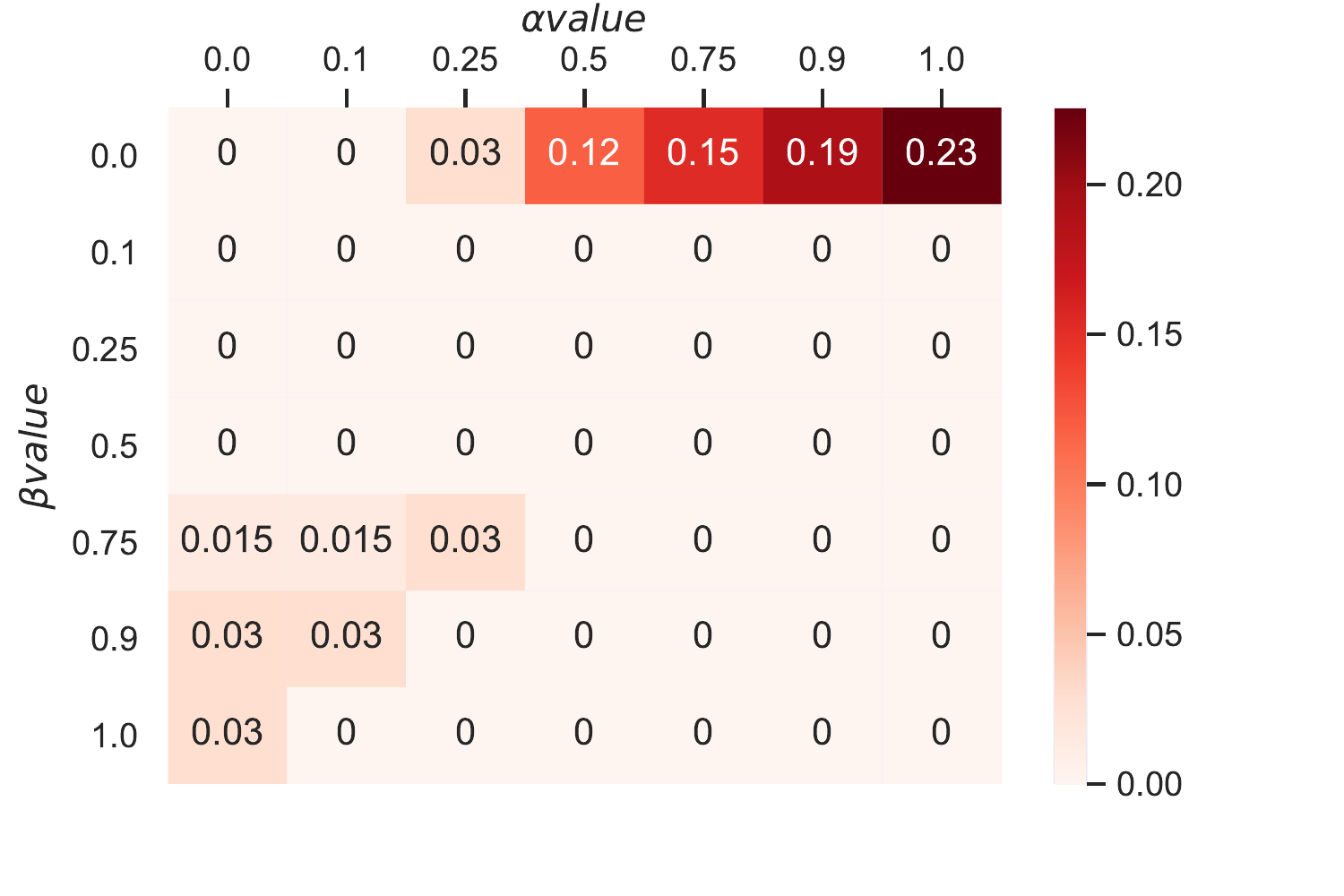} & \includegraphics[width=0.3\linewidth]{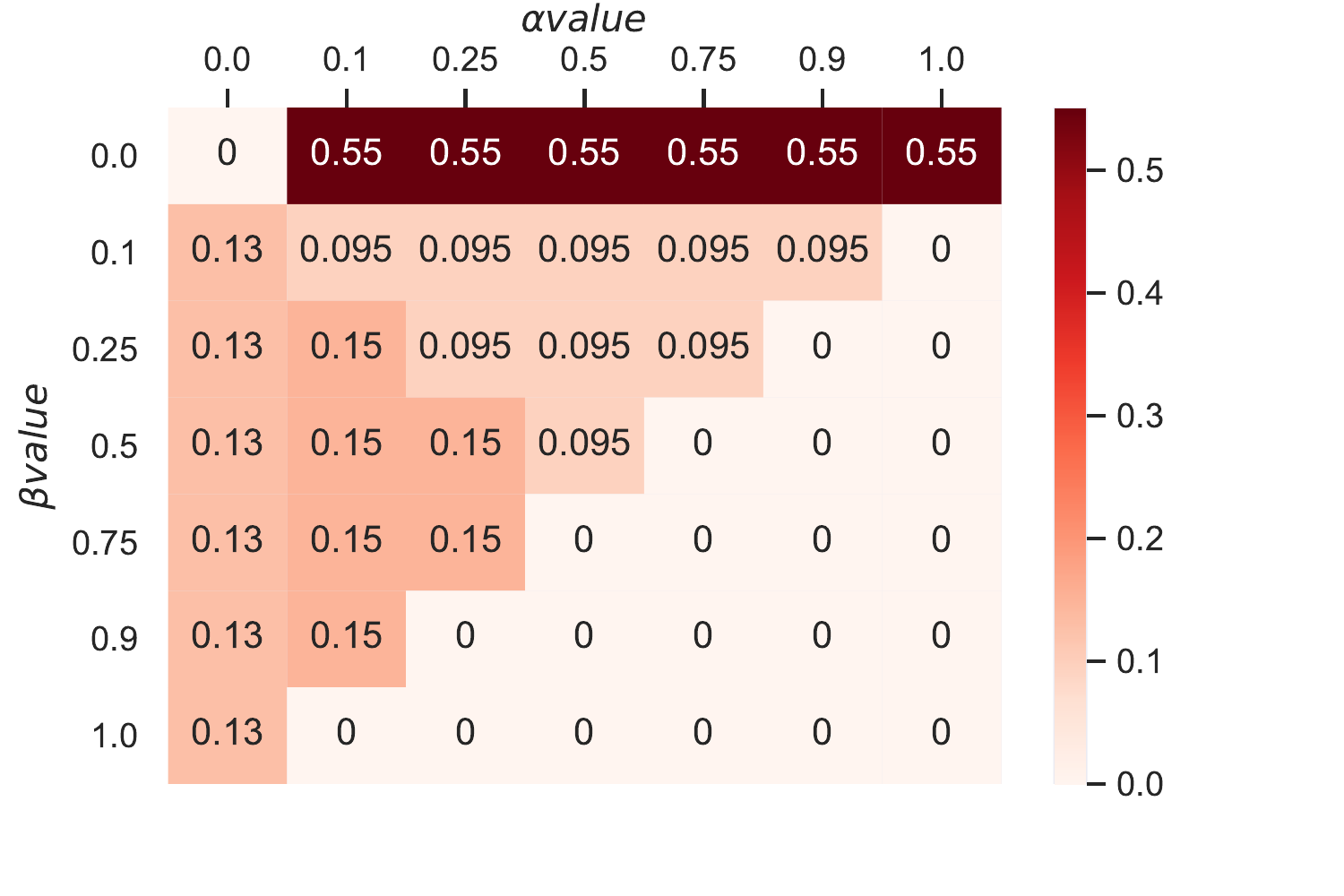} &
         \includegraphics[width=0.3\linewidth]{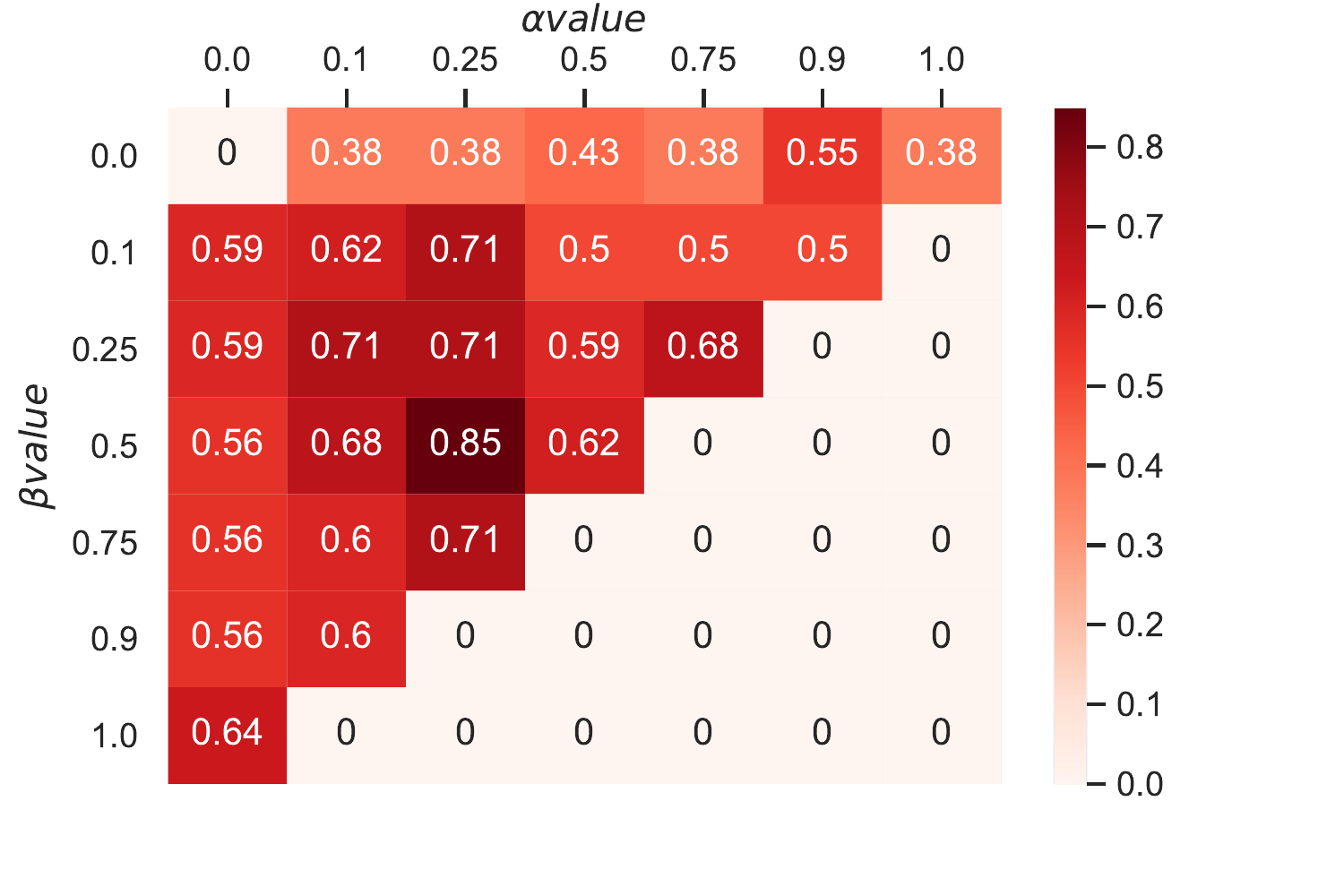}  \\
         \midrule
         Dolphin & \includegraphics[width=0.3\linewidth]{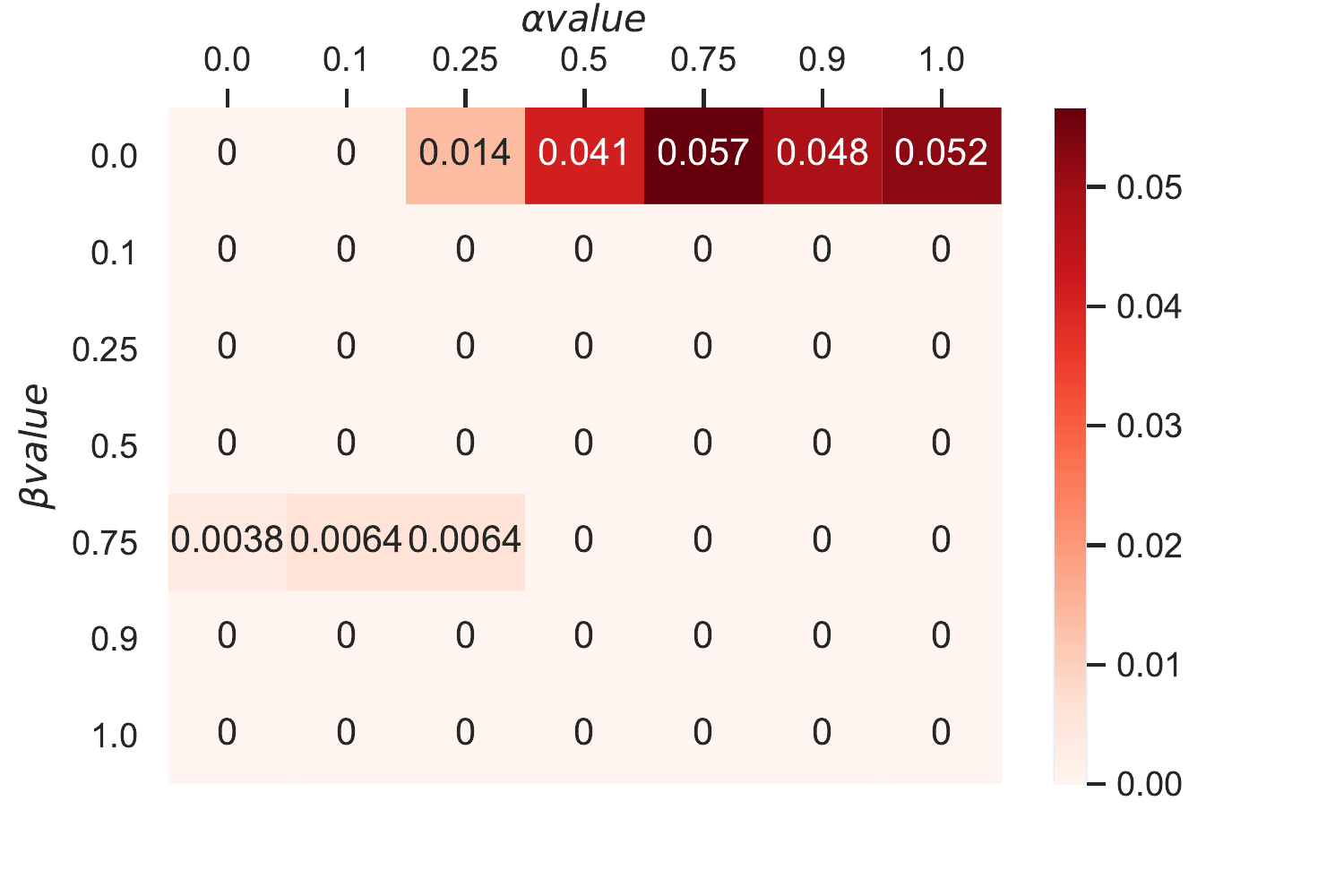} & \includegraphics[width=0.3\linewidth]{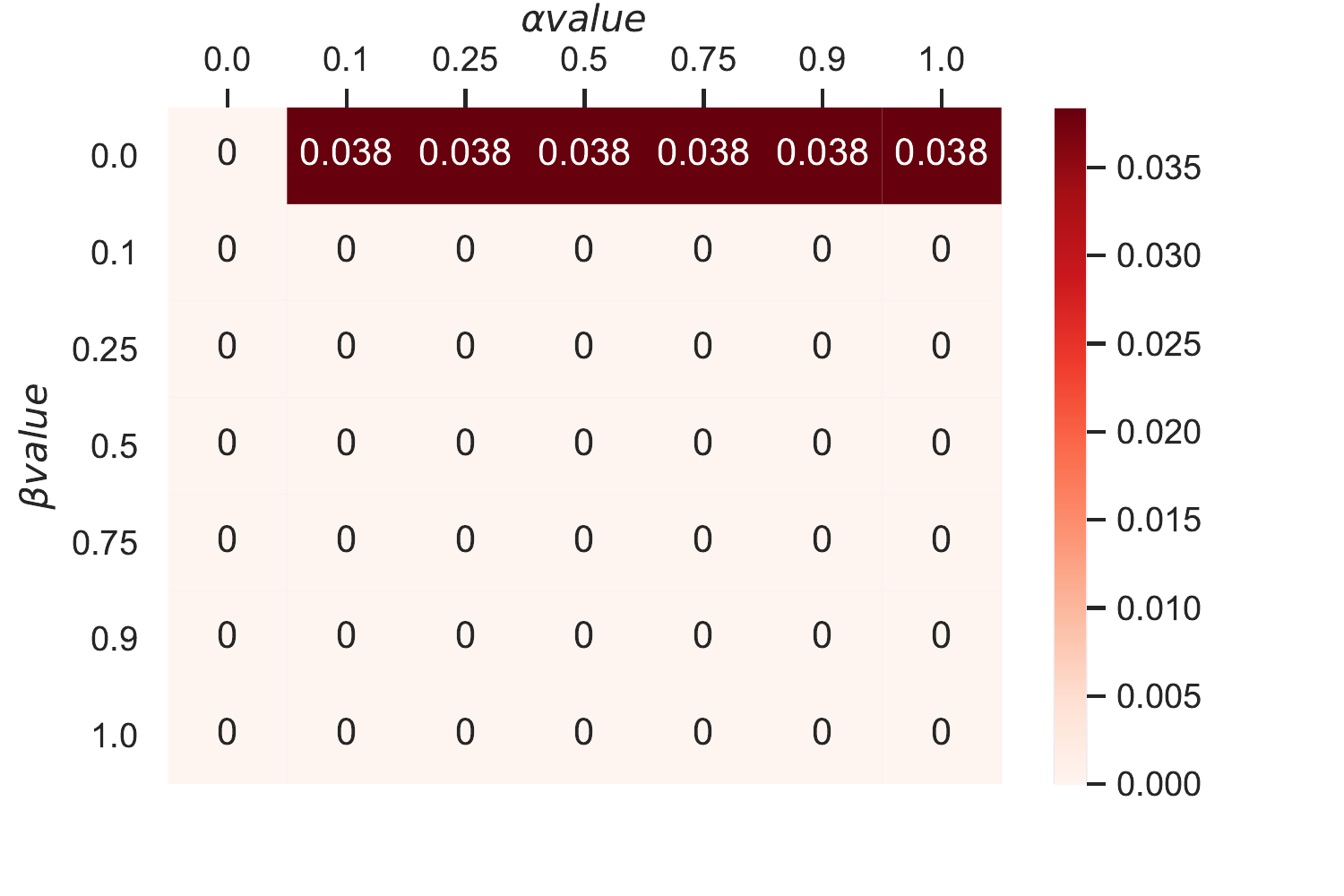} & \includegraphics[width=0.3\linewidth]{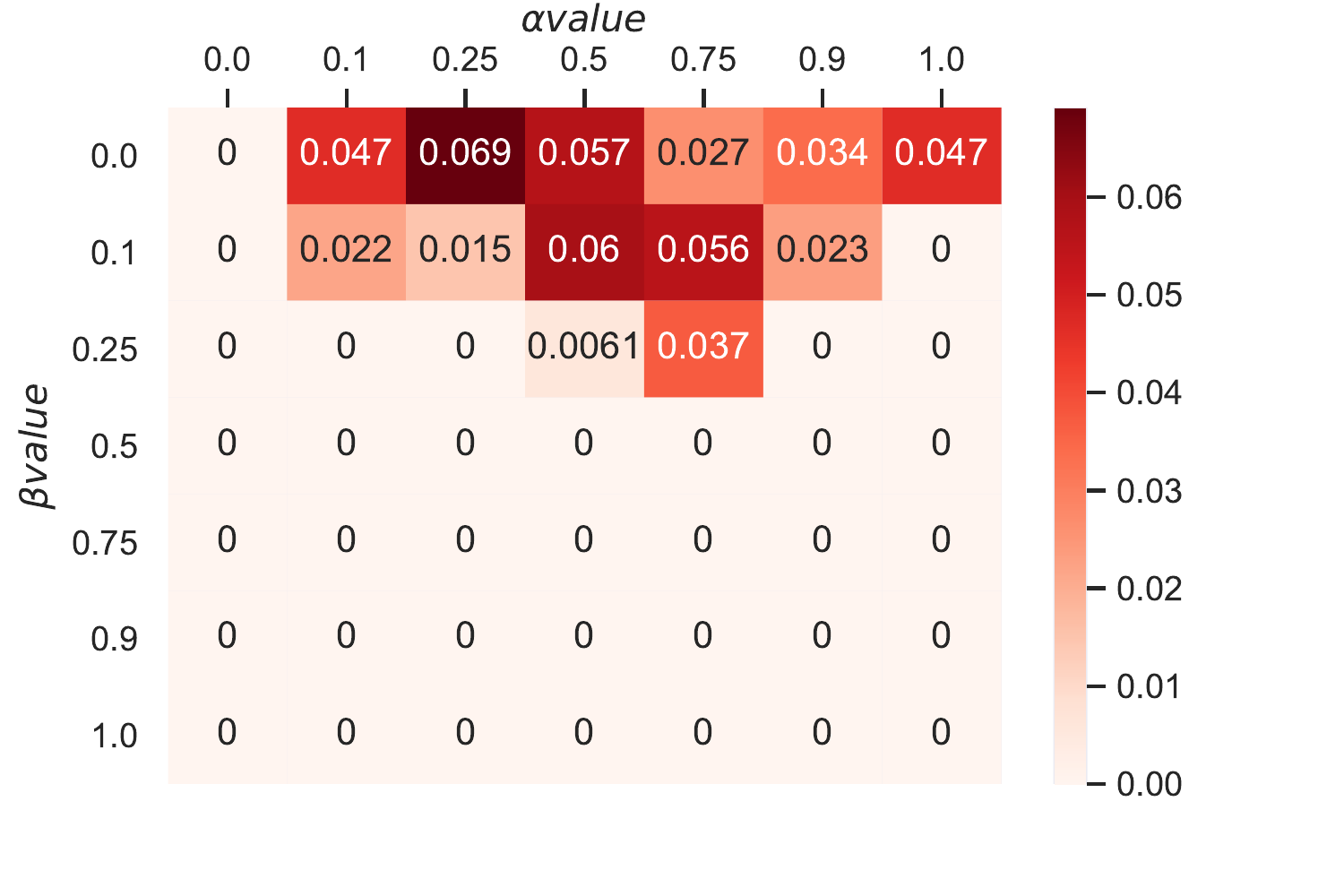}  \\
         \midrule
         Polbooks & \includegraphics[width=0.3\linewidth]{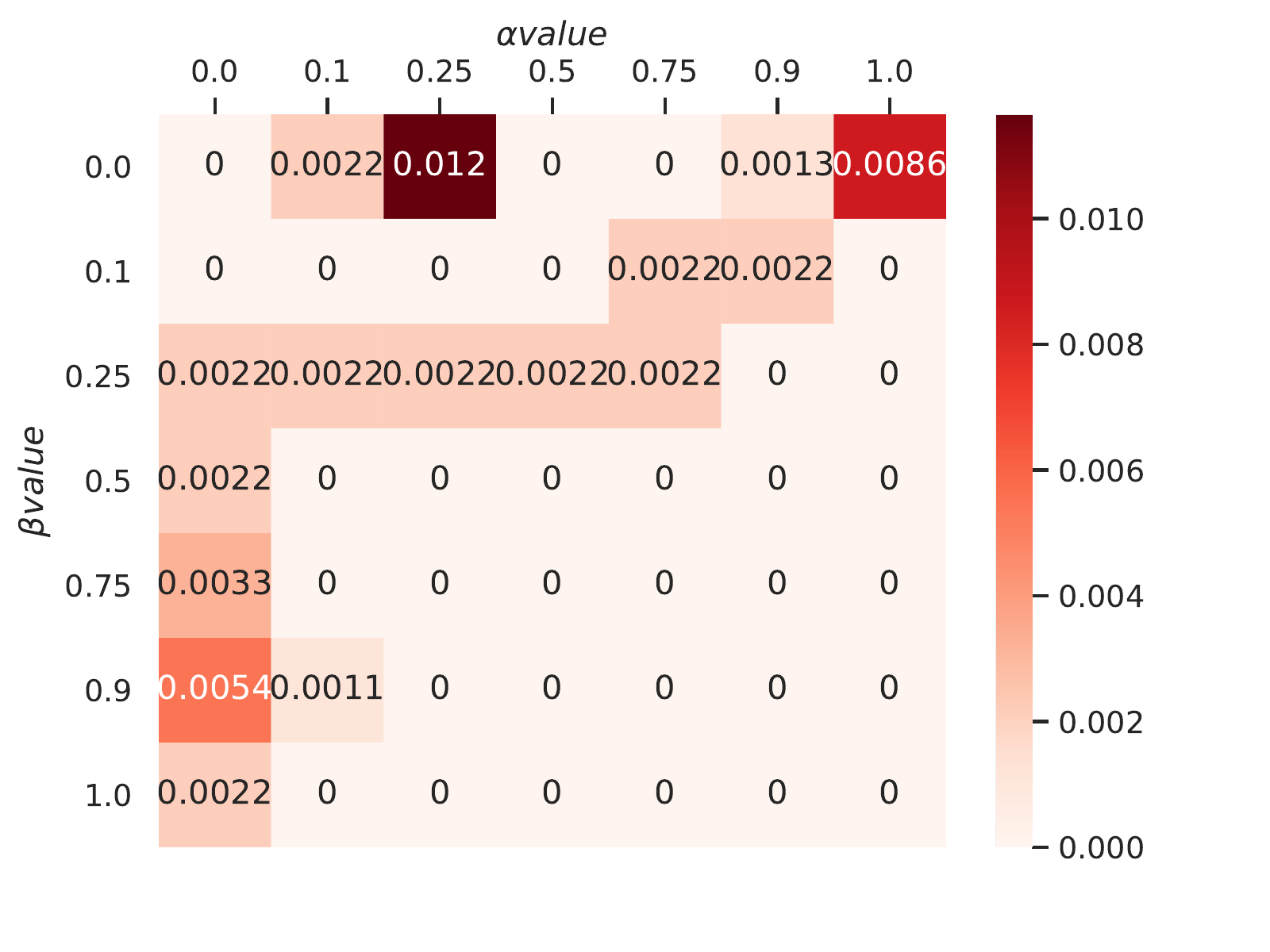} & \includegraphics[width=0.3\linewidth]{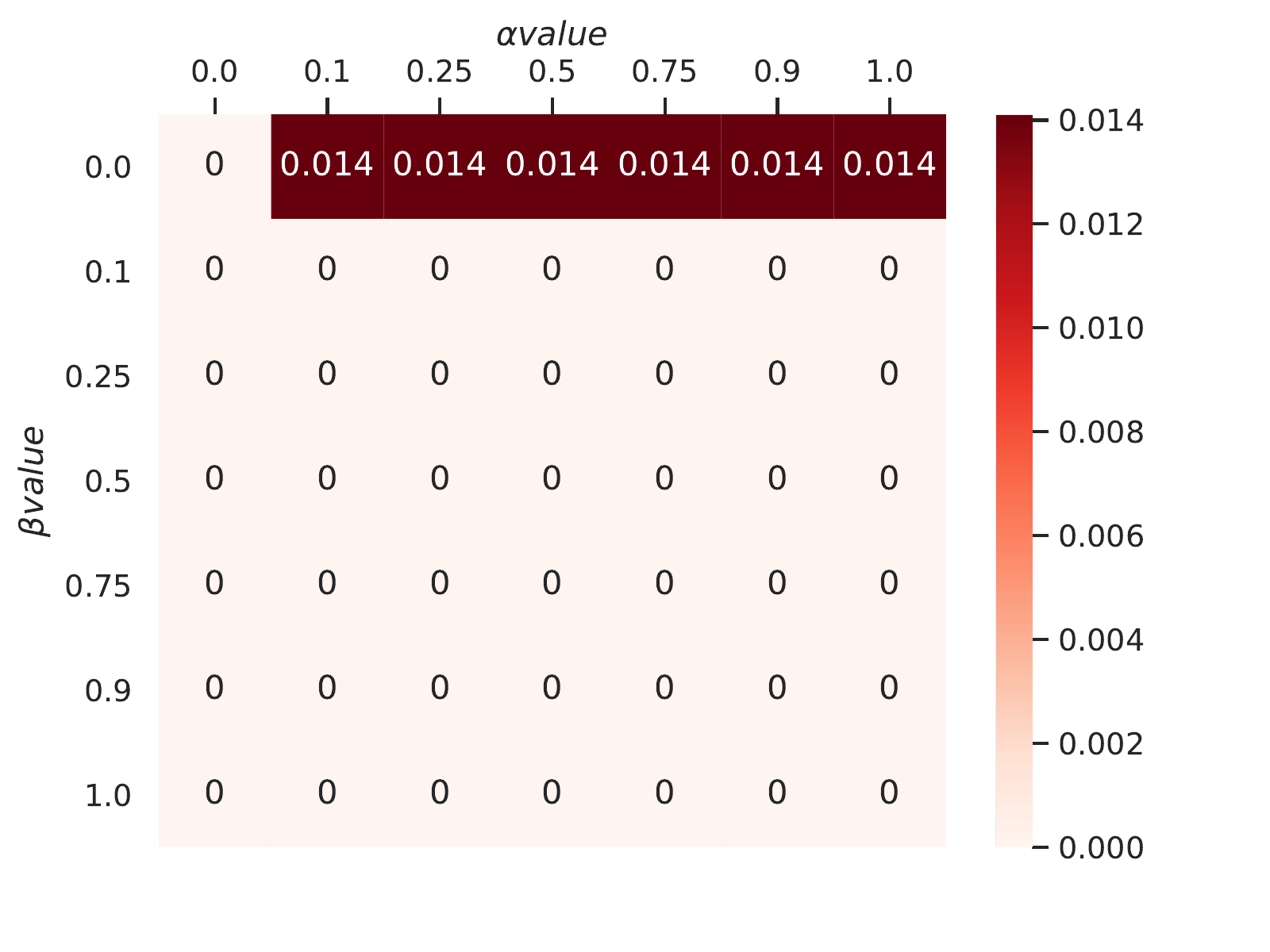} & \includegraphics[width=0.3\linewidth]{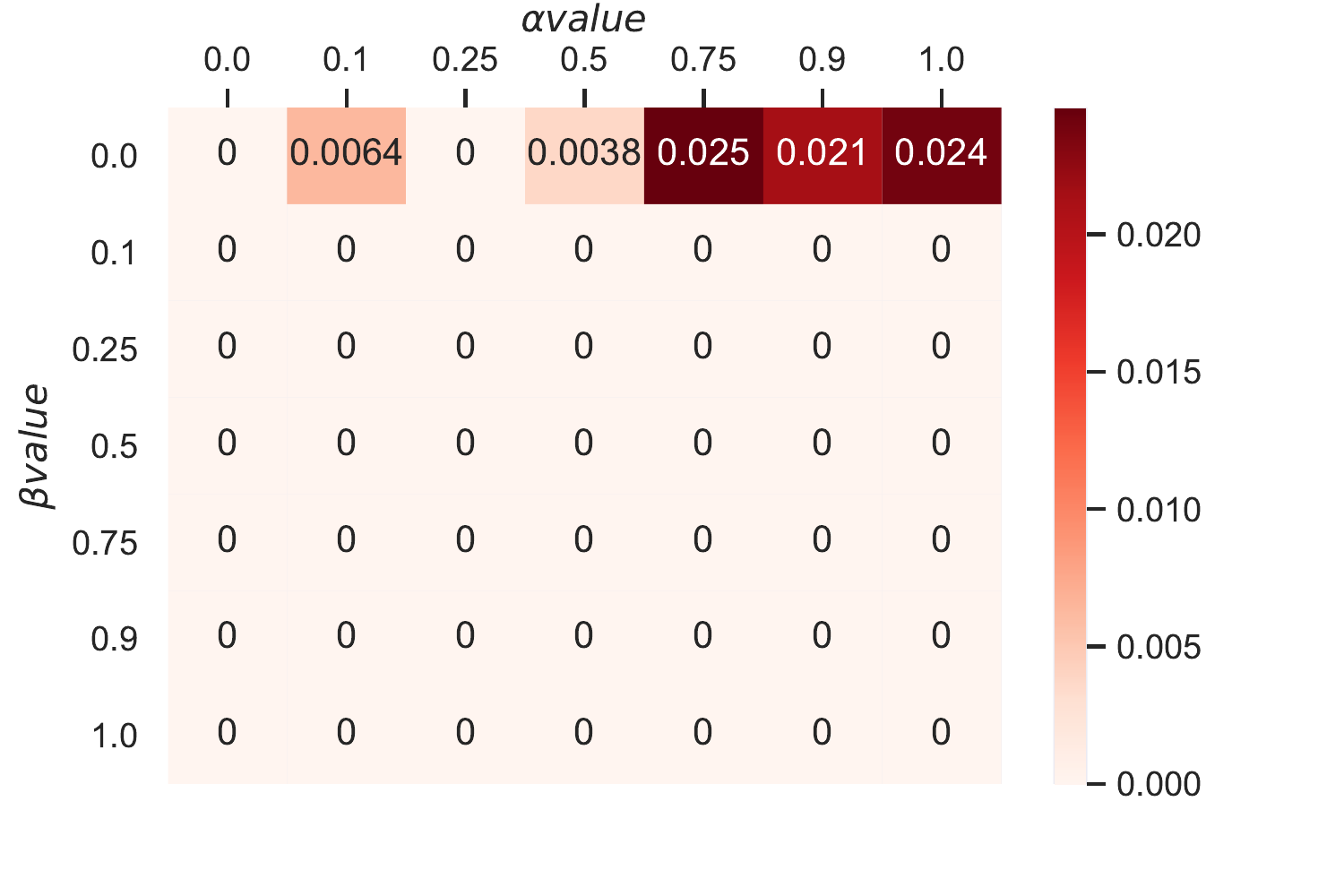}  \\
         \bottomrule
    \end{tabular}
    \caption{NMI results for the community detection algorithms usgin affinity networks. We found good results for the Zachary karate network dataset, but underwhelming results for the Dolphin and Polbooks in all cases.}
    \label{fig:nmi}
\end{figure*}

We have studied how these indexes change when we apply the three algorithms using the affinity matrix resulting when using the combinations proposed in Eq. (\ref{eq:ncc}) instead of the original adjacency matrix. We have reported the modularity results in Figure \ref{fig:mod} for the modularity index, and in Figure \ref{fig:nmi} for the NMI index.

A summary of the best results can be found in Table \ref{tab:nmi_best}, alongside the results of the Borgia Clustering for the same trials. We found that in most cases the best friend affinity obtains the best results, specially in the case of the Greedy Modularity and the Girvan-Newman algorithms. However, in the case of the Louvain algorithm, non-convex combinations of affinity functions resulted best for the NMI index. Compared to the results obtained using the adjacency matrix, NMI index decreased dramatically for the Dolphin and Polbooks datasets, but modularity seemed to improve in most cases, specially for the Louvain algorithm.

All the studied algorithms seemed to obtain good modularity results, both in the case of the adjacency and the affinity matrices. In fact, these algorithms seemed to over perform the Borgia Clustering in terms of the modularity index, although the Borgia Clustering presented much higher NMI values than the rest of the algorithms, probably because the Borgia Clustering computed correctly the number of real communities for each dataset.

\begin{table}[ht]
    \centering
    \begin{tabular}{cccc}
    \toprule
    Dataset & Algorithm &  NMI & Mod. \\
    \midrule
        & Greedy Mod. & 0.45 & 0.38\\
        Zachary & Girvan-Newman & 0.35 & 0.41\\
        & Lovaine &  0.35 & 0.41\\
        \midrule
        & Greedy Mod. & 0.41 & 0.49 \\
        Dolphin & Girvan-Newman &  0.44 & 0.51 \\
        & Lovaine & 0.33 &  0.51\\
        \midrule
        & Greedy Mod. & 0.49 & 0.50 \\
        Polbooks & Girvan-Newman & 0.48 & 0.51\\
        & Lovain & 0.56 & 0.49\\
        \bottomrule
    \end{tabular}
    \caption{Results for community detection for the three datasets considered using three different algorithms. In this case we used the adjacency matrix to compute the network partition. }
    \label{tab:classic}
\end{table}

\begin{table}[ht]
    \centering
    \begin{tabular}{cccc}
    \toprule
    Dataset & Algorithm & NMI &  Mod. \\
    \midrule
            &  Greedy Mod. &  0.31 & 0.23\\
    Zachary &  Girvan Newman & 0.42 & 0.55 \\
            &  Louvain & 0.31 & 0.85\\
            & Borgia Clustering & 0.83 & 0.36\\
            \midrule
         &  Greedy Mod. & 0.052 & 0.43 \\
    Dolphin &  Girvan Newman & 0.038 & 0.55\\
            &  Louvain & 0.089 &  0.59 \\
            & Borgia Clustering & 1.00 & 0.37\\
            \midrule
            &  Greedy Mod. & 0.086 & 0.55 \\
    Polbooks &  Girvan Newman & 0.014 & 0.55 \\
            &  Louvain & 0.024 & 0.59\\
            & Borgia Clustering & 0.56 & 0.49\\
            \bottomrule
    \end{tabular}
    \caption{A summary for the best results for community detection computed over affinity networks for the datasets and algorithms studied. }
    \label{tab:nmi_best}
\end{table}

\section{Results and Discussion} \label{sec:res}

In this experimentation we studied the performance of different community detection algorithms using less than-convex combinations of affinity functions, in three different datasets. We found satisfactory results for the modularity metric in all datasets for all the algorithms tested, and for the two metrics studied in the Zachary karate network. Also, we found the best friend affinity to be the most effective affinity function to perform community detection. 

The Louvain algorithm seems to be the most successful classical algorithm when using affinity functions, according to the modularity index. The Girvan-Newman algorithm also obtained a moderate performance increase. On the contrary, the Greedy modularity algorithm performed better with the original adjacency networks compared to the affinity functions tested. None of the studied algorithms performed better than the Borgia Clustering in terms of real world labels, but they did in terms of modularity.

A possible explanation for the disparity between the modularity and NMI indexes performance could be due to the social interactions exploited by the affinity functions, which could be different from the dynamics registered by the ground truth labels. Besides, the behaviour of some of these algorithms can be affected dramatically by the affinity function chosen. For example, the best common friend affinity leads to an increase in the number of edges in the network. This will for sure affect the performance and execution time of the Girvan-Newman algorithm, that needs to recompute the edge betweenness in each iteration.

\section{Conclusions} \label{sec:conclusions}

In this work we have proposed a new method to construct affinity functions using non-convex combinations of affinity functions. We have explained how these combinations work and how some centrality measures change based on the mixing factors of the combination. Besides, we have used the affinity functions for the first time in classical community detection algorithms designed to work with the adjacency matrix. In order to test our proposal, we have computed a series of experiments using different combinations of affinity functions, and we have compared the best results obtained with the classical algorithms with the Borgia Clustering. 

We found that the use of affinity functions can have a positive impact on the modularity metric performance, although evaluation with ground truth labels decreased in most cases. However, in the case of the Borgia Clustering we found the opposite results. The ground truth labels evaluation was in general terms better than the rest of the algorithms but the modularity values were lower.

Future research shall study the relationship between the Borgia Clustering and the rest of the community detection algorithms. As the Borgia Clustering is very expensive to compute in large networks, we are particularly interested in a possible hybrid algorithm that combines the Louvaine algorithm and the Borgia Clustering, in order to surpass some of the limitations of the latter in terms of memory and execution time. Also, we shall research the interaction of affinity functions with specific community detection algorithms in order to develop specific functions to enhance the performance for each individual algorithm.
\section{Acknowledgments}
Javier Fumanal Idocin and Humberto Bustince’s re-search has been supported by the project PID2019-108392GBI00 (AEI/10.13039/501100011033).\\
Maria Minárová research has been funded by the project work was supported by the projects APVV-17-0066 and APVV-18-0052. \\
Oscar Cordón's research was supported by the Spanish Ministry of Science, Innovation and Universities under grant EXASOCO (PGC2018-101216-B-I00), including, European Regional Development Funds (ERDF).


\bibliographystyle{ieeetr}
\bibliography{socialbib}

\end{document}